\documentclass[preprint,review,12pt]{elsarticle}

\usepackage{graphicx}
\usepackage{amssymb}
\usepackage{amsmath}
\usepackage{IEEEtrantools}
\usepackage{amsthm}
\usepackage{float}
\biboptions{square,comma,sort&compress}
\usepackage[letterpaper, margin=1in]{geometry} 
\usepackage{adjustbox}
\usepackage{multirow}

\journal{}

\begin{document}

\begin{frontmatter}


\title{Comprehensive Energy Balance Analysis of Photon-Enhanced Thermionic Power Generation Considering Concentrated Solar Absorption Distribution}% Force line breaks with \\



\author{A. N. M. Taufiq Elahi$^1$} 
\author{Mohammad Ghashami$^2$}

\author{Devon Jensen$^3$}
\author{Keunhan Park$^{1*}$}

\address{$^{1}$Department of Mechanical Engineering, University of Utah, Salt Lake City, UT 84112, USA}
\address{$^{2}$Department of Mechanical and Materials Engineering, University of Nebraska-Lincoln, Lincoln, NE 68588, USA}
\address{$^{3}$Advanced Cooling Technologies, Inc., Lancaster, PA 17601, USA}
\begin{abstract}
The present article reports a comprehensive energy balance analysis of a photon-enhanced thermionic emission (PETE) device when it is used for concentrated solar power (CSP) generation. To this end, we consider a realistic PETE device composed of a boron-doped silicon emitter on glass and a phosphorus-doped diamond collector on tungsten separated by the inter-electrode vacuum gap. Depth-dependent spectral solar absorption and its photovoltaic and photothermal energy conversion processes are rigorously calculated to predict the PETE power output and energy conversion efficiency. Our calculation predicts that when optimized, the power output of the considered PETE device can reach 1.6 W/cm$^2$ with the energy conversion efficiency of $\sim$18 \% for 100$\times$ solar concentration, which is substantially lower than those predicted in previous works under ideal conditions. In addition, the photon-enhancement ratio is lower than 10 and decreases with the increasing solar concentration due to the photothermal heating of the emitter, suggesting that PETE may be an adequate energy conversion process for low-to-medium CSP below $\sim$100$\times$ concentration. These observations signify the importance of a rigorous energy balance analysis based on spectral and spatial solar absorption distribution for the accurate prediction of PETE power generation. 
\end{abstract}

\begin{keyword}
Photon-Enhanced Thermionic Emission (PETE) \sep Concentrated Solar Power (CSP) \sep Thermionic Energy Conversion (TEC) \sep Energy Balance Analysis 


\end{keyword}

\end{frontmatter}

\bigskip

%\linenumbers

\section{Introduction}
\label{S:1}

A thermionic energy converter (TEC) is a solid-state heat engine that directly generates electric power from heat by thermionic emission of electrons \cite{Hatsopoulos1973,Hatsopoulos1979}. A vacuum TEC consists of emitter and collector electrodes separated by a vacuum gap, such that electrons vaporized from a hot emitter can be collected by a relatively cold collector. When the two electrodes are connected by an electrical load, electric current flows through the load due to the potential difference to generate electric power. TEC is beneficial because of its pollution-free and noise-free power generation. However, a TEC device needs a very high operating temperature and yet suffers from low energy conversion efficiency due to a high vacuum energy barrier \cite{Hatsopoulos1973,Hatsopoulos1979}. To mitigate these challenges, \citet{Schwede2010} proposed photon-enhanced thermionic emission (PETE) that utilizes photoexcitation of electrons for thermionic power generation by replacing a metallic emitter with a $p$-doped semiconductor. When the $p$-doped semiconductor emitter is illuminated by photons having energies above the band gap, photoexcitation of electrons increases the electron population in the conduction band, reducing the effective energy barrier for electron emission \cite{nelson2003physics,Green2003333}. In addition, thermalization of excessive photon energy increases the emitter temperature to facilitate thermionic emission of electrons. As a result, PETE is expected to generate power with a much higher energy conversion efficiency than a conventional TEC even at lower operating temperatures \cite{Schwede2010}. 
%The electron emission process in a PETE device is dictated by the illumination of the high-energy photons, which are absorbed to photoexcite electrons and increase the emitter temperature by thermalization. 

The photon-enhancement of thermionic emission has been verified experimentally for different semiconducting materials  \cite{schwede2013photon,sun2014thermally,Tang2014,Zhuravlev2014,Zhuravlev2017}. Theoretical models have also been developed to understand the underlying physics of the PETE process and applied to predict the performance of a PETE device for concentrated solar power (CSP) \cite{Varpula2012a,Sahasrabuddhe2012,varpula2015,YANG2015410,FENG2018,yang2020temperature}. However, they handled the emitter temperature as an independent input variable without considering the energy balance of the system. Several works have conducted the energy balance analysis of the PETE device to calculate the dependence of the emitter temperature on incident solar radiation and its effect on the performance of PETE power generation \cite{Segev2012,Segev2013b, Su2013,su2014material, Segev2015,su2016photon,Kribus2016,sandovsky2016,xiao2018thermodynamic}. However, the obtained results are not consistent to each other, revealing that the accuracy of the energy balance analysis stringently depends on the adequate modeling of charge and thermal transports due to photothermal interactions of solar energy with electrode materials. Lack of rigor in the modeling of transport processes based on unrealistic material properties could lead to an overprediction of the PETE performance.

In the present article, we report the comprehensive energy balance analysis of a PETE device when concentrated solar radiation is incident onto the device. Instead of assuming perfect solar absorption by the thermionic emitter \cite{Segev2012,Segev2013b,Su2013,Segev2015,Segev2015a,sandovsky2016,Wang2016,WANG2019}, real material properties are implemented to calculate the locally distributed spectral solar absorption in the emitter structure. 
%The emitter structure under consideration has a thin semiconductor layer for thermionic emission on top of a thick glass substrate for structural support of the emitter. The glass substrate also allows the transmission of the visible to near-infrared solar spectrum to photo-excite electrons in the emitter. 
Energy balance analysis is rigorously applied to the emitter structure to calculate its temperature distribution upon the illumination of solar radiation with different concentration factors. For accurate calculation of charge transport for a wide range of the interelectrode vacuum gap distance, we take into account the negative space charge effect \cite{ito2012optically,SU2014137,Reck2014a,Segev2015a,Buencuerpo2015,Kribus2016} and the image charge effect \cite{Wang2016} to the potential barrier profile. In addition to the energy transfer by electrons, radiative heat transfer should be considered as another energy transport mechanism between the electrodes. Particularly when the interelectrode vacuum gap is set within the single-digit micrometer or smaller range, a PETE device would suffer from significant radiative heat loss due to the near-field interactions of thermally induced evanescent electromagnetic (EM) waves in the vacuum space \cite{Park2013,liu2019effects,WANG2019}. Near-field radiative heat loss from the emitter structure to the collector is calculated by implementing the multi-layer dyadic Green's function within the fluctuational electrodynamics framework \cite{Park2008c}. Our study also presents optimal design and operating parameters for photon-enhanced thermionic concentrated solar power generation, along with the expected power output and the energy conversion efficiency.

\section{Modeling}
\label{S:2}
\begin{figure}
\centering\includegraphics[width=0.8\linewidth]{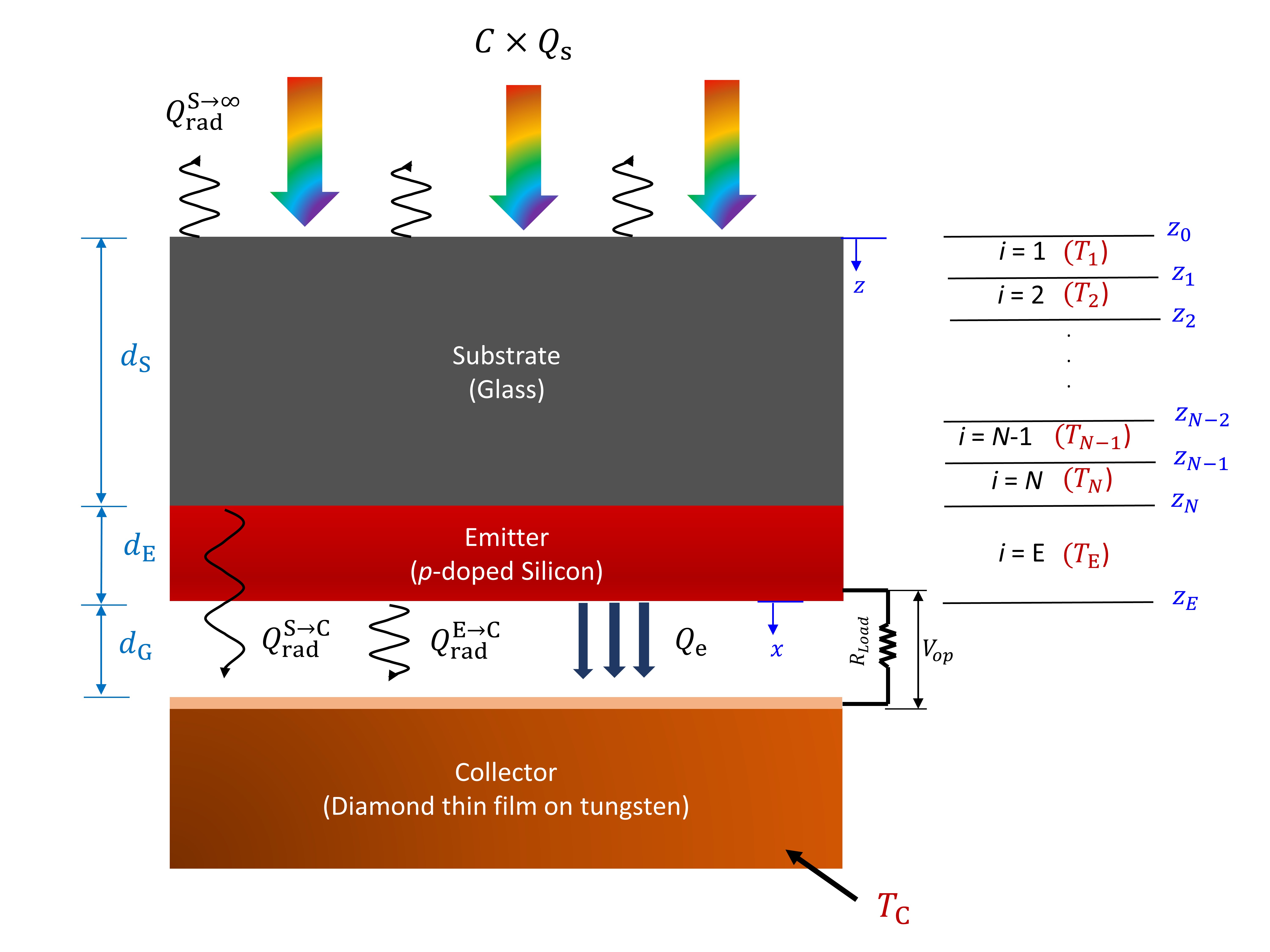}
\caption{}
\label{Fig1}
\end{figure}

Fig. \ref{Fig1} illustrates the schematic diagram of the modeled PETE device under concentrated solar irradiation. The emitter electrode is configured as a boron-doped single crystalline silicon (Si) layer on a soda-lime glass substrate. Si is selected as an emitter material in the present study because it can withstand a higher operational temperature compared to other candidate materials, such as GaAs and InP, which start decomposing above 870 K for GaAs and above 750 K for InP \cite{varpula2015}. The glass substrate is required to structurally supports the Si emitter layer while transmitting high-energy photons to the Si emitter for photoexcitation of electrons. In addition, it absorbs some low-energy photons below the bandgap of Si, which is beneficially used to heat the emitter for thermalization. On the other hand, a phosphorus-doped diamond film deposited on tungsten by chemical vapor deposition (CVD) is considered as a thermionic collector due to its lowest work function measured to date (i.e., 0.9 eV) \cite{Koeck2009}. %When the PETE device is connected to an electrical load ($R_\mathrm{load}$), thermionically emitted electrons are swept to the collector due to the potential difference, which leads to the generation of electric power at the load. 
%Concentrated solar energy is utilized as input energy for this device. Here, high energy photons of the solar spectrum help in photoexcitation of electron when the other part of the spectrum is used for heat-induced electron emission.

For the energy balance analysis of the PETE device upon the concentrated solar irradiation, the emitter structure is discretized into $N$ layers for the glass substrate, each at temperature $T_{i}$ ($i=1,2,3 ... N$), and the Si emitter layer at $T_\mathrm{E}$ ($i=\mathrm{E}$). When the collector temperature is assumed to be uniform at $T_\mathrm{C}$, the energy balance equation for each layer of the emitter structure can be written as
\begin{subequations}
    \label{Eq:energy_S}
    \begin{IEEEeqnarray}{rCl}
        Q^{1}_\mathrm{abs}- Q^\mathrm{ S\rightarrow\infty}_\mathrm{rad}-Q^{1\rightarrow2}_\mathrm{con}&= 0& ~~~\mathrm{for}~~ i=1 \\
       %\vspace{6pt}
        %\nonumber \\ 
         Q^{i}_\mathrm{abs}+Q^{(i-1)\rightarrow i}_\mathrm{con}-Q^{i\rightarrow (i+1)}_\mathrm{con} &= 0& ~
         ~~\mathrm{for}~~ 2\le i \le N-1 \\
        %\vspace{6pt}
        %\nonumber \\ 
        Q^{N}_\mathrm{abs}+Q^{(N-1)\rightarrow N}_\mathrm{con}-Q^{N\rightarrow \mathrm{E}}_\mathrm{con}- Q^\mathrm{ S\rightarrow C}_\mathrm{rad} &= 0& ~~~\mathrm{for}~~ i = N
    \end{IEEEeqnarray}
\end{subequations}
for the glass substrate and
    \begin{equation}
        Q^\mathrm{E}_\mathrm{abs}+Q^{N\rightarrow \mathrm{E}}_\mathrm{con}- Q^\mathrm{ E\rightarrow C}_\mathrm{rad}- Q_e = 0 
    \label{Eq:energy_E}
    \end{equation}
for the Si emitter. Here, $Q^{i}_\mathrm{abs}$ is the total solar irradiance absorbed in the $i$-th layer, $Q_\mathrm{rad}$ is the radiative heat flux, $Q^{i\rightarrow (i+1)}_\mathrm{con}$ is the conduction heat flux from the $i$-th to ($i+1$)-th layer, and $Q_e$ is the net energy flux carried by electrons. It should be noted that the convection heat loss from the top surface of the glass substrate is ignored by assuming that the PETE device is vacuum-packaged. The solar absorption per unit area in the $i$-th layer ($Q^{i}_\mathrm{abs}$) can be calculated by $Q^{i}_\mathrm{abs} = \int_0^\infty [G_\lambda(z_{i-1},\omega)-G_\lambda(z_{i},\omega)]d\omega$, where $G_\lambda (z_i,\omega)$ is the spectral solar irraidance (or flux) at interface $z_i$ and $\omega=2\pi c_0/\lambda$ is the angular frequency with $c_0$ and $\lambda$ being the speed of light in free space and the wavelength of solar radiation, respectively. In the present study, the spectral solar irradiance at each interface of the multi-layered emitter structure is calculated by the scattering-matrix method \cite{ Francoeur2009ay,Deparis:11}. 

Radiative heat loss is another key feature for the accurate energy balance analysis. Radiative heat loss from the glass substrate to the environment is expressed as $Q^\mathrm{ S\rightarrow\infty}_\mathrm{rad}=\varepsilon \sigma( {T^4_1}-{T^4_\infty})$, where $\varepsilon$ is the total emissivity of glass, $\sigma$ is the Stefan-Boltzmann constant, and $T_\infty$ is the surrounding temperature \cite{gubareff1960thermal}. Near-field radiative heat losses from the glass substrate and the Si emitter to the collector are calculated within the fluctuational electrodynamics framework:   
\begin{subequations}
    \label{Eq:energy_S}
    \begin{IEEEeqnarray}{rCl}
    Q^\mathrm{ S\rightarrow C}_\mathrm{rad}=\int_\mathrm{0}^\infty {S^\mathrm{ S\rightarrow C}_z(T_N,\omega)} d\omega \\
    \nonumber \\
    Q^\mathrm{ E\rightarrow C}_\mathrm{rad}=\int_\mathrm{0}^\infty {S^\mathrm{ E\rightarrow C}_z(T_\mathrm{E},\omega)} d\omega,
\end{IEEEeqnarray}
\end{subequations}
where superscripts S, E, and C represent the glass substrate, the emitter, and the collector, respsctively, and $S_\mathrm{z}$ is the time-averaged Poynting vector in the $z$-direction between layers as indicated by the arrow, which can be calculated using the multi-layer dyadic Green's function \cite{Park2008c}. For the calculation of the near-field radiative heat transfer, we assume that the collector is composed of a thin CVD diamond film on top of a semi-infinite tungsten at the uniform temperature, $T_\mathrm{C}$. In addition, $S_z^\mathrm{ S\rightarrow C}$ is calculated by assuming that the glass substrate is semi-infinite at $T_N$, which is an adequate assumption due to a short penetration depth of near-field thermal radiation \cite{Basu2009c}. 
The conduction heat flux from the $i$-th to ($i+1$)-th layer in the emitter structure can be approximated as $Q^{i\rightarrow (i+1)}_\mathrm{con}= k_\mathrm{eff} \left(T_{i}-T_{i+1}\right)/\Delta z$, where $k_\mathrm{eff}$ is the effective thermal conductivity and $\Delta z$ is the distance between the center points of the $i$-th and ($i+1$)-th layers \cite{patankar2018numerical}. For the present study, dielectric functions were taken from literature for soda-lime glass \cite{RUBIN1985275}, boron-doped Si \cite{Fu2006a}, CVD diamond \cite{Dore:98}, and tungsten \cite{Lee2012j}. Among them, the temperature dependences of dielectric functions were considered for boron-doped Si and tungsten. The temperature-dependent thermal conductivities were also found from Ref. \cite{sergeev1982thermophysical} for soda-lime glass and Ref. \cite{Shank1963} for Si. It should be noted that the thermal conductivity of intrinsic Si is used for the present study due to the lack of available data for $p-$doped Si at high temperatures. However, we believe that this approximation would not significantly alter the result as the conduction thermal resistance of the Si emitter is three orders of magnitude smaller than that of the glass substrate. 

The net energy flux transported by electrons from the emitter to the collector is expressed as \cite{Ghashami2017,WANG2019}
\begin{equation} 
    \label{Eq:Qe}
    Q_{e}=\frac{1}{q}\left[(J_\mathrm{E}-J_\mathrm{C})W_\mathrm{max} + 2k_\mathrm{B}(J_\mathrm{E}T_\mathrm{E}-J_\mathrm{C}T_\mathrm{C})\right],
\end{equation}
where $J_\mathrm{E(C)}$ is the current density of the emitter (collector), $W_\mathrm{max}$ is the maximum potential barrier, $k_\mathrm{B}$ is the Boltzmann constant, and $q$ is the electron charge. It should be noted that a portion of the electron-carried energy is used to generate electric power while the remainder is lost as heat. 
%It should be noted that Eq. (\ref{Eq_Qe}) formulates the net transfer of potential energy and the kinetic energy of electrons .  
Solar radiation whose energy is greater than the energy bandgap of Si photoexcites electrons to the conduction band, which greatly enhances the concentration of electrons in the conduction band ($n$) over the equilibrium value ($n_\mathrm{eq}$). The thermionic current density from the emitter due to photoexcited electrons can be written as \cite{Wang2016}
\begin{equation} 
\label{Eq:J_E}
    J_\mathrm{E}=\left(\frac{n}{n_\mathrm{eq}}\right)A_\mathrm{E} T_\mathrm{E}^2  \exp\left(-\frac{W_\mathrm{max}}{k_\mathrm{B} T_\mathrm{E}}\right),
\end{equation}
where $A_\mathrm{E}=120~\mathrm{A/cm^2K^2}$ is the Richardson constant of the Si emitter \cite{Schwede2010}. It should be noted that $J_\mathrm{E}^0 = A_\mathrm{E} T_\mathrm{E}^2  \exp(-W_\mathrm{max}/k_\mathrm{B} T_\mathrm{E})$ denotes the conventional thermionic current density, indicating that $n/n_\mathrm{eq}$ can be defined as \textit{the (photon-)enhancement ratio}. Similarly, the collector current density is expressed as $J_\mathrm{C}=A_\mathrm{C} T_\mathrm{C}^2  \exp[-({W_\mathrm{max}}-qV_\mathrm{op})/{k_\mathrm{B} T_\mathrm{C}}]$, where $A_\mathrm{C}=120~\mathrm{A/cm^2K^2}$ is the Richardson constant of the diamond collector \cite{Koeck2009,Schwede2010} and $V_\mathrm{op}$ is the operational voltage drop across the electrical load. 

The carrier concentration in the conduction band of the Si emitter under photoexcitation ($n$) is calculated by the following equation \cite{Schwede2010,Ghashami2017}:
\begin{equation} 
\label{Eq:Gamma_Balance}
    \Gamma_\mathrm{gen}-\Gamma_\mathrm{rec}-\Gamma_e=0,
\end{equation}
which balances the photogeneration rate of electron-hole pairs due to solar absorption ($\Gamma_\mathrm{gen}$), the recombination rate of electron-hole pairs ($\Gamma_\mathrm{rec}$), and the thermionic emission rate of photoexcited electrons from the emitter ($\Gamma_e$). The photogeneration rate of electron-hole pairs can be expressed as $\Gamma_\mathrm{gen}=(1/d_\mathrm{E})\int_{\hbar\omega\geq E_\mathrm{g}} \left[Q_\mathrm{abs,\lambda}^\mathrm{E}(\omega)/\hbar\omega\right] d\omega$, 
%\begin{equation} 
%    \label{Eq:Gamma_gen}
%    \Gamma_\mathrm{gen}=\frac{1}{d_\mathrm{E}}\int_{{hc}/{\lambda} \geq E_\mathrm{g}} \frac{ Q_\mathrm{abs}^\mathrm{E}(\lambda)}{{hc}/{\lambda}}  d\lambda,
%\end{equation}
where $d_\mathrm{E}$ is the emitter thickness, $\hbar$ is the reduced Plank constant, $E_\mathrm{g}$ is the energy bandgap of the emitter, and $Q_\mathrm{abs,\lambda}^\mathrm{E}$ is the spectral absorption of the solar irradiance in the emitter. 
%and can be calculated by $Q_\mathrm{abs,\lambda}^\mathrm{E}=G_\lambda(z_{N+1},\omega)-G_\lambda(z_\mathrm{E},\omega)$. 
In the present study, the temperature-dependence of the Si bandgap is considered by $E_\mathrm{g}=1.170-(4.73\times 10^{-4})T_\mathrm{E}^2/(T_\mathrm{E}+636)$ eV \cite{sze2006physics}. The recombination rate ($\Gamma_\mathrm{rec}$) is determined by considering the near-field radiative recombination ($\Gamma_\mathrm{rad}$), Auger recombination ($\Gamma_\mathrm{Aug}$), Shockley-Read-Hall recombination ($\Gamma_\mathrm{SRH}$) and surface recombination ($\Gamma_\mathrm{surf}$) mechanisms, i.e., $\Gamma_\mathrm{rec}=\Gamma_\mathrm{rad}+\Gamma_\mathrm{Aug}+\Gamma_\mathrm{SRH}+\Gamma_\mathrm{surf}$. Detailed formulation of each type of recombination can be found in other works \cite{Segev2013b,Ghashami2017} and will not be repeated here. Finally, the thermionic emission rate of photoexcited electrons ($\Gamma_e$) can be calculated by $\Gamma_e=\left(J_\mathrm{E}-J_\mathrm{E}^0\right)/qd_\mathrm{E}$.

$W_\mathrm{max}$ can be determined from the potential energy barrier profile $W(x)$ in the inter-electrode gap space. $W(x)$ is written as
\begin{equation} 
\label{eq12}
    W(x) = W_\mathrm{\mathrm{sc}}(x) +  W_\mathrm{\mathrm{ic}}(x) 
\end{equation}
based on the potential barrier profile due to electron accumulation in the inter-electrode space, $W_\mathrm{\mathrm{sc}}(x)$, and the image charge perturbation, $W_\mathrm{\mathrm{ic}}(x)$. When electrons accumulate in the inter-electrode vacuum space and build up negative space charges, it hinders further emission of electrons from the emitter surface. Under the assumption that electrons traveling in the inter-electrode gap are collisionless, the space charge potential can be calculated by numerically solving Poisson's equation \cite{Langmuir1923,Hatsopoulos1973,Hatsopoulos1979,WANG2019}:
\begin{equation} 
\label{Eq:Poisson}
\frac{d^2W_\mathrm{sc}(x)}{dx^2}=-\frac{q^2N_\mathrm{e}(x)}{\epsilon_{0}},
\end{equation}
where $x$ is the location between the emitter ($x= 0$) and the collector $(x=d_\mathrm{G})$, $N_e(x)$ is the local electron number density, and $\epsilon_{0}$ is the permittivity of vacuum. Boundary conditions for Eq. (\ref{Eq:Poisson}) are $W_\mathrm{sc}(0)=\Phi_\mathrm{E}$ and $W_\mathrm{sc}(d_\mathrm{G})=\Phi_\mathrm{C}$ while $dW_\mathrm{sc}/dx=0$ at $x=x_\mathrm{max}$ ($0\leq x_\mathrm{max} \leq d_\mathrm{G}$), where $\Phi_\mathrm{E(C)}$ is the work function of the emitter (collector). For a semiconductor emitter, the work function can be expressed as $\Phi_\mathrm{E}=\chi+E_g-E_F$, where $\chi$ and $E_F$ are the electron affinity and the Fermi energy of the emitter, respectively. The numerical framework to calculate $W_\mathrm{sc}$ has been discussed in detail in Ref. \cite{Hatsopoulos1973,Hatsopoulos1979,WANG2019}. 
%If there is no space charge accumulation in the vacuum space, the potential barrier becomes ideal with a linear shape and can be written as\cite{Simmons1963a,Baldea2012a}:
%\begin{equation} \label{eq13}
%  W_\mathrm{\mathrm{id}}(x) = \Phi_\mathrm{E} - (\Phi_\mathrm{E}-\Phi_\mathrm{C}-qV_\mathrm{op})\left(\frac{x}{d_\mathrm{G}}\right)
%\end{equation}
%where  $\Phi_\mathrm{E(C)}$ is the work functions of the emitter (collector). The work function for a semiconductor emitter can be written in terms of band gap ($E_\mathrm{g}$), Fermi level($E_\mathrm{F}$), and electron affinity($\chi$),  $\Phi_\mathrm{E}=E_\mathrm{g}-E_\mathrm{F}+\chi$. Now, we can consider the space charge potential as the difference between the potential barrier with and without space charge accumulation, $W_\mathrm{sc}(x)=W_\mathrm{ie}(x)-W_\mathrm{id}(x)$. 
On the other hand, the image charge potential $W_{\mathrm{ic}}(x)$ is calculated by considering the electrostatic forces between image charges formed in the electrodes and the electrons in the vacuum gap \cite{Simmons1963a,Hishinuma2001,Baldea2012a,Wang2016}:
\begin{equation} 
\label{eq15}
  W_{\mathrm{ic}}(x) = \frac{q^2}{16\pi \epsilon_0 d_\mathrm{G}}\left[-2\Psi(1)+\Psi\left( \frac{x}{d_\mathrm{G}}\right)+\Psi\left(1-\frac{x}{d_\mathrm{G}}\right)\right]
\end{equation}
where $\Psi$ is the digamma function. It should be noted that the image charge effect competes with the space charge effect in modifying the potential barrier. For a larger inter-electrode vacuum gap, the space charge effect plays a dominant role to increase the potential barrier. However, as the inter-electrode gap distance decreases, the image charge effect becomes important to lower the potential barrier while the space charge effect is suppressed due to the sweep of electrons under the stronger electric field between the electrodes \cite{Jensen2017b}. 
 
Since charge and thermal transport processes are strongly coupled, the energy balance analysis should be carefully conducted through iterations. After the initial calculation of $Q_\mathrm{abs}^i$ in each layer, the temperature distribution of the emitter structure is calculated iteratively until the incoming and outgoing energy fluxes of the Si emitter in Eq. (\ref{Eq:energy_E}) are balanced within the convergence criterion of 0.001\%. From the energy balance analysis and the thermionic current density calculation, the electric power density can be determined by
\begin{equation}
    P_\mathrm{net}=(J_\mathrm{E}-J_\mathrm{C})V_\mathrm{op},
    \label{Eq:P_net}
\end{equation}
%where $V_\mathrm{op}$ is the operational voltage that generates the maximum power from the PETE device. 
In addition, the energy conversion efficiency is defined as 
\begin{equation}
    \eta=\frac{P_\mathrm{net}}{C\times Q_s},
\end{equation}
where $C$ is the concentration factor and $Q_s$ is the total solar irradiance with the global standard spectrum (AM1.5D). Therefore, when the concentration factor is fixed, the energy conversion efficiency should be linearly proportional to the generated power density.
Table \ref{table:1} summarizes input and geometrical parameters that are optimized through the energy balance analysis. More discussions about the optimization process will be provided in the following section. Unless otherwise mentioned, these parameters are used to calculate the PETE power generation. The collector temperature ($T_\mathrm{C}$) is set to 540 K by satisfying the optimal collector condition, i.e., $T_\mathrm{C}=600\times\Phi_\mathrm{C}$ \cite{Hatsopoulos1973,Lee2012j,WANG2019}. The effect of the collector temperature onto the PETE performance is further discussed in Fig. \ref{Fig7}.

\section{Results and Discussion}
\label{S:3}
\begin{figure}%[hbt!]
\centering\includegraphics[width=0.8\linewidth]{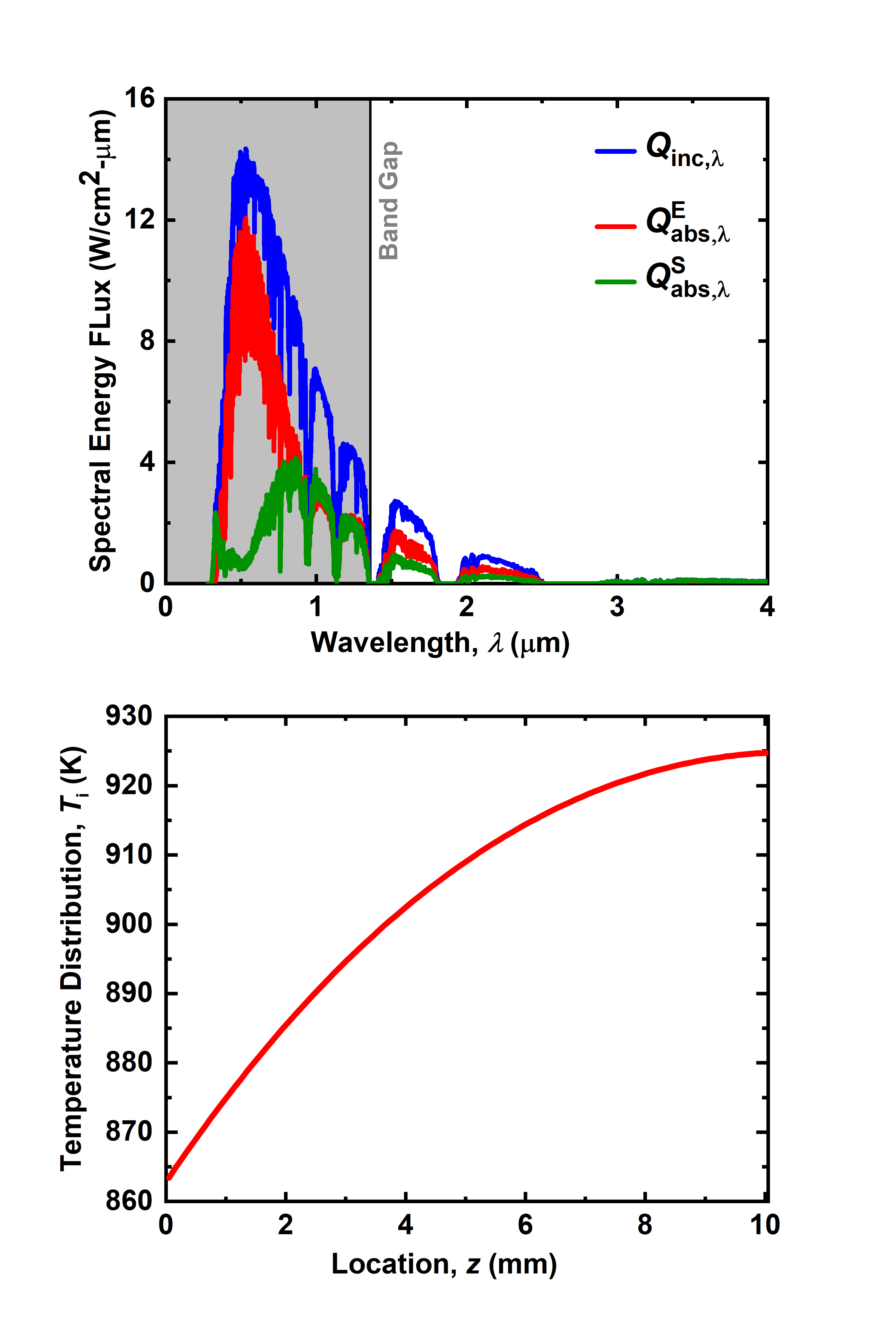}
\caption{}
\label{Fig2}
\end{figure}

When concentrated solar radiation is incident on the PETE device, part of the solar energy is absorbed as it interacts with each of the layers. Fig. \ref{Fig2}(a) shows the solar absorption spectra in the glass substrate ($Q_\mathrm{abs,\lambda}^\mathrm{S}$) and the Si emitter layer ($Q_\mathrm{abs, \lambda}^\mathrm{E}$). In the figure, the shaded region denotes the above-bandgap spectrum of Si at 925 K, which is the calculated temperature of the Si emitter at the default operation conditions (Table \ref{table:1}). It should be noted that the temperature dependence of the Si bandgap was considered here (i.e., $E_g =$ 0.91 eV at 925 K) as described in the Modeling section. For this particular case, 57.2 \% of the above-bandgap solar energy is absorbed by the Si emitter to photoexcite electrons, while 25.3 \% of the above-bandgap energy is absorbed by the glass substrate. In addition, sub-bandgap energy is also absorbed by the Si emitter (50.5\%) and the glass substrate (32.6 \%). All absorbed solar energy not used for photoexcitation is converted to heat the emitter structure, which is in fact beneficial for thermalization of electrons. Fig. \ref{Fig2}(b) shows the resultant local temperature profile of the emitter structure. The top surface of the emitter assembly ($z=0$) has a lower temperature (864 K) due to radiative heat loss to the environment. While strong absorption of sub-bandgap energy in the glass substrate is desired to increase the emitter layer temperature for thermionic emission, it also induces substantial radiavie heat loss to the environment. Although the present study considers glass as a readily available and cost-effective substrate material, an optically engineered substrate can be implemented to maximize thermionic emission by making a good balance between photoexcitation and thermalization of electrons in the emitter. 

\begin{figure}%[H]
\centering
\includegraphics[width=0.8\linewidth]{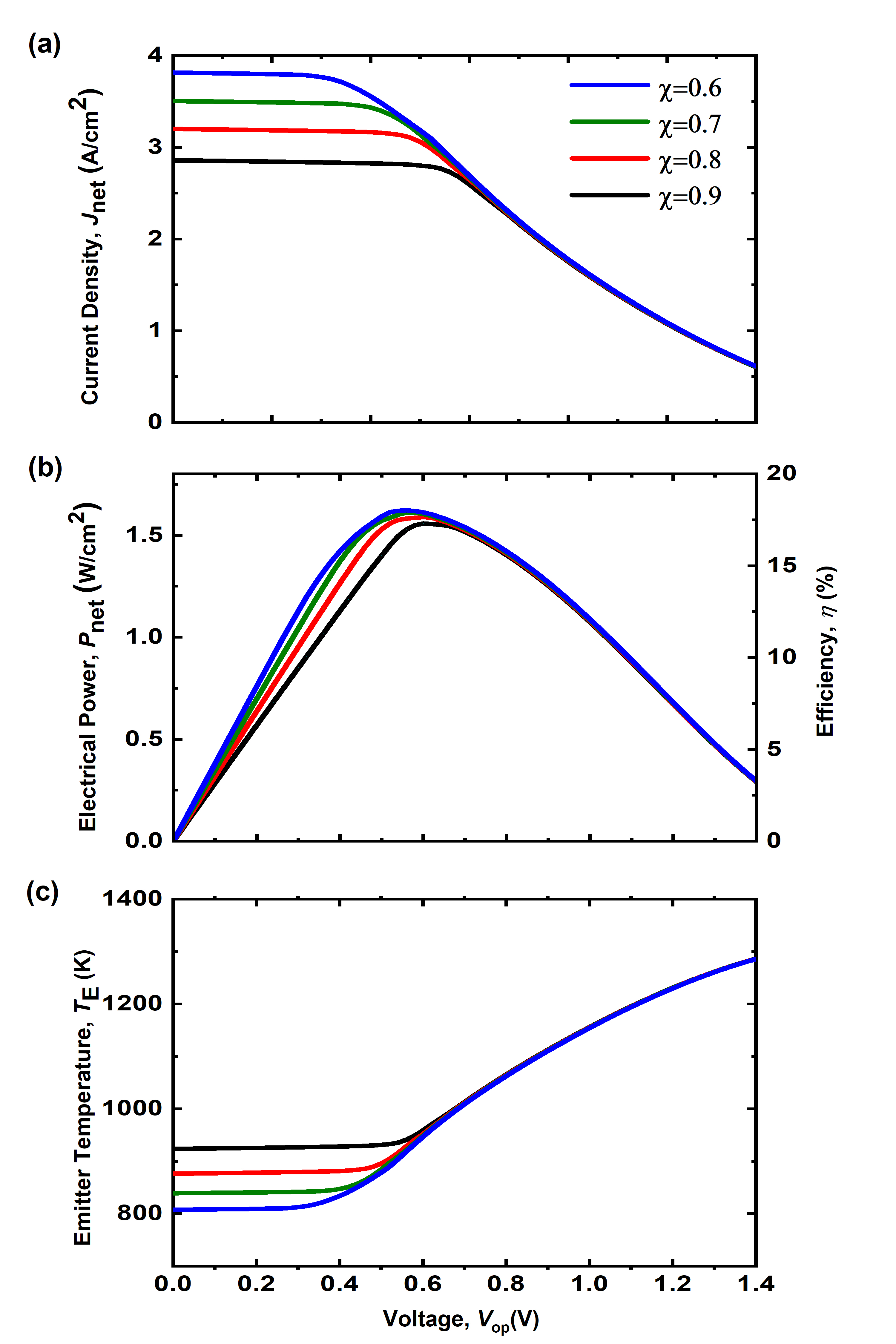}
\caption{}
\label{Fig3}
\end{figure}

The net PETE current density ($J_\mathrm{net}=J_\mathrm{E}-J_\mathrm{C}$) as a function of the operational voltage is shown in Fig. \ref{Fig3}(a) for different electron affinities of Si. Other design parameters still remain the same as provided in Table \ref{table:1}. The general trend of the calculated $J-V$ curve is similar to the conventional thermionic emission, which shows a nearly constant current density up to the flat-band voltage ($V_\mathrm{FB}=(\Phi_\mathrm{E}-\Phi_\mathrm{C})/q$) and decays in the Boltzmann regime ($V_\mathrm{op}>V_\mathrm{FB}$) \cite{Hatsopoulos1979}. As a result, the PETE power density and the efficiency reach the maxima at the optimal operating voltage around $V_\mathrm{FB}$: see Fig. \ref{Fig3}(b). However, the PETE power density ($P_\mathrm{net}$) and the efficiency ($\eta$) will be kept and used to denote the maximum values at the optimal operating voltage for simplicity in the remaining results and discussion. Figs. \ref{Fig3}(a) and (b) also show the effect of the electron affinity to the PETE current and the power densities. Previous studies have demonstrated that the electron affinity of Si can be tuned over a wide range even to a negative value by cesium deposition or oxygen-potassium co-adsorption \cite{goldstein1973leed,martinelli1974thermionic,morikawa1995further,smestad2004conversion}. The obtained results demonstrate that PETE power generation is benefited by lowing the electron affinity of the emitter. A low electron affinity leads to a low potential barrier within the inter-electrode vacuum space, allowing more electron emission from the emitter in the flat-band regime. However, the maximum power output (the efficiency) only increases from 1.56 W/cm$^2$ (17.3 \%) to 1.62 W/cm$^2$ (18.0 \%) when the electron affinity changes from 0.9 eV to 0.6 eV. Particularly, the power density curves for $\chi=0.6$ eV and $0.7$ eV are almost identical, suggesting that the electron affinity of the Si emitter should be optimized at 0.7 eV for the best cost-effectiveness. The electron affinity also affects the emitter temperature as can be seen in Fig. \ref{Fig3}(c). A low electron affinity facilitates the emission of hot electrons by lowing the potential barrier, resulting in a decrease of the emitter temperature. When optimized at $\chi=0.7$ eV, the emitter temperature at the maximum power output is calculated to be 925 K, which is much lower than an operational temperature of conventional thermionic power generation. 

\begin{figure}%[H]
\centering
\includegraphics[width=0.8\linewidth]{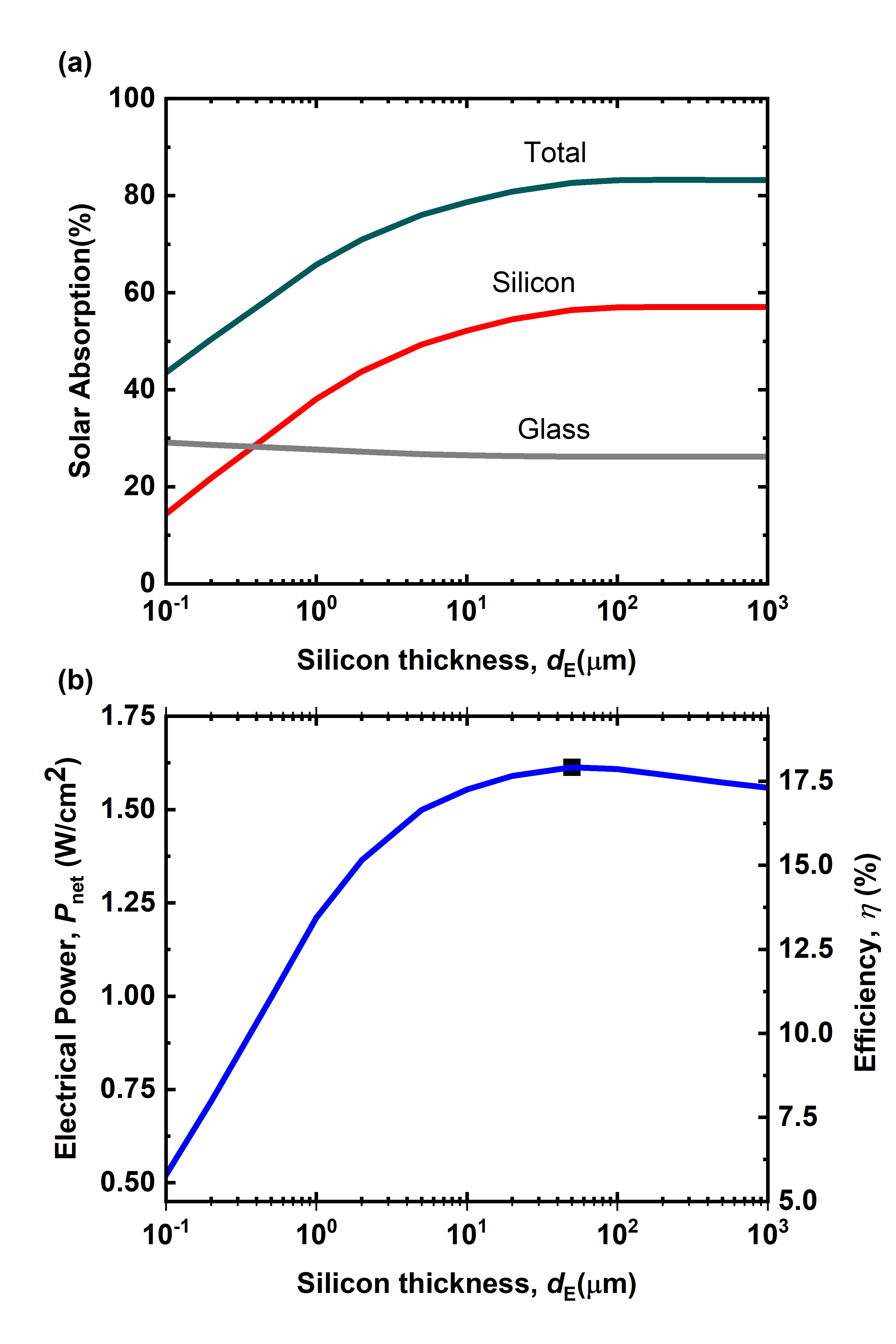}
\caption{}
\label{Fig4}
\end{figure}

The thickness of the semiconducting emitter layer may be one of the most important factors that govern PETE power generation. As shown in Fig.\ref{Fig4}(a), almost $57\%$ of the total incident solar radiation can be absorbed by the Si layer when its thickness becomes $\sim 50$ $\mu$m. However, further increase of the Si layer thickness does not help increase the solar absorption. On the other hand, the glass substrate absorbs $\sim$26\% of the incident solar radiation, although its value is slightly affected by the Si layer thickness. For a sub-micron Si layer, solar radiation is reflected back from the bottom surface of the Si layer and absorbed by the substrate again. The remaining $\sim17\%$ is reflected from the top surface of the glass substrate, which is considered as the optical loss. Fig. \ref{Fig4}(b) shows the PETE power output as a function of the Si layer thickness. The increasing electric power output for $d_\mathrm{E} \lesssim 50$ $\mu$m is consistent with the solar absorption trend in the Si emitter, indicating that photoexcited electrons drive thermionic power generation. However, the power output begins to slightly decrease with the further increase of the Si layer thickness because more electron-hole pairs start recombining than being photogenerated as the emitter becomes thicker than the optimal value \cite{YANG2015410}. Therefore, the Si emitter thickness is optimized to $50$ $\mu$m that suppresses the recombination of electron-hole pairs while the solar absorption by the Si emitter is fully secured. 

\begin{figure}%[H]
\centering
\includegraphics[width=0.8\linewidth]{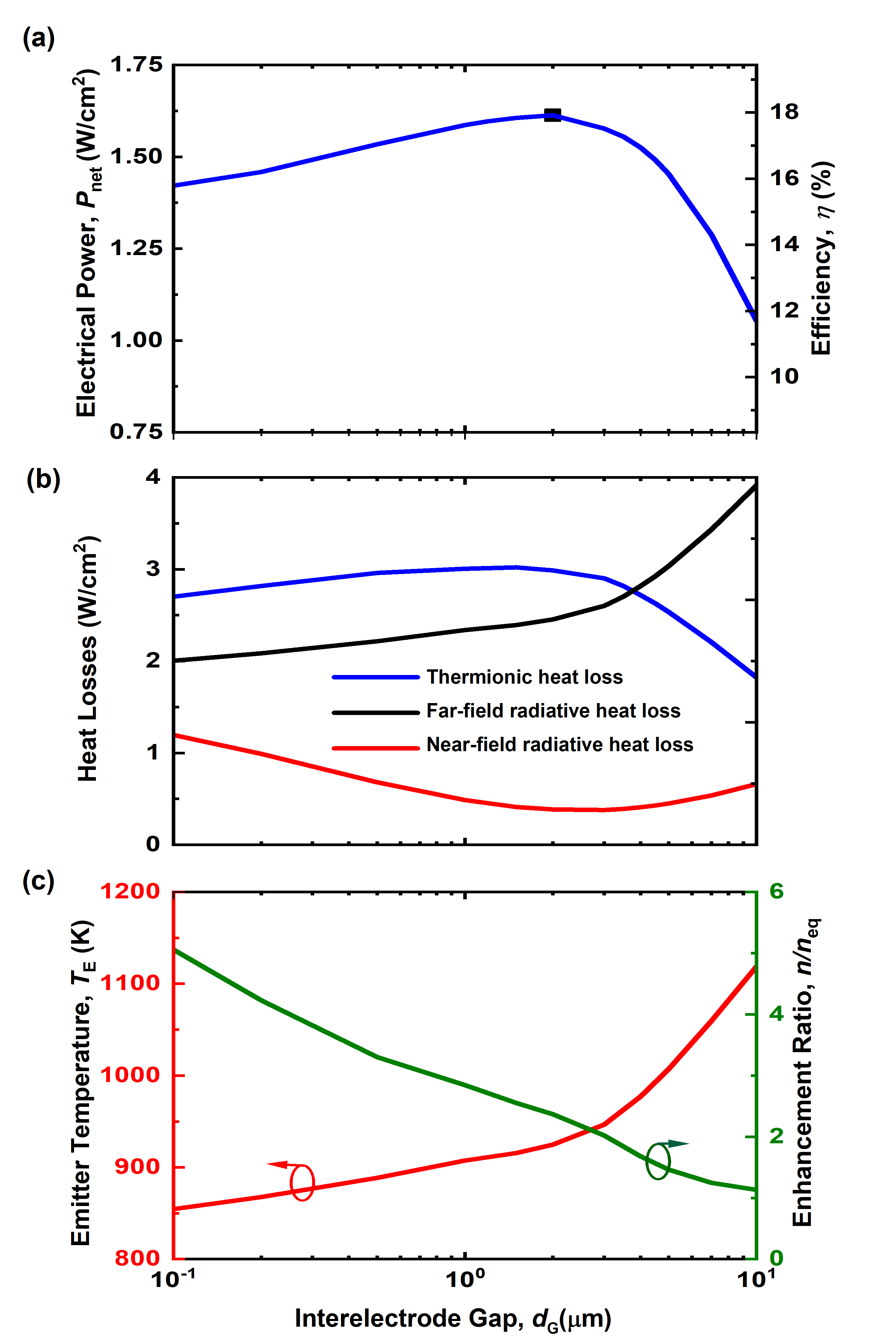}
\caption{}
\label{Fig5}
\end{figure}

Another key parameter for PETE power generation is the inter-electrode vacuum gap distance. The gap-dependent PETE behaviors shown in Fig. \ref{Fig5} demonstrate strongly coupled charge and thermal transport processes across the vacuum gap.  For large gap distances above 2 $\mu$m, the negative charge effect plays a dominant role in the PETE process. The buildup of the potential barrier (or $W_\mathrm{max}$) as a result of the negative charge effect inhibits thermionic emission of electrons, which is manifested by the drastic decrease of the PETE power density ($P_\mathrm{net}$) and the thermionic heat loss (i.e., $Q_\mathrm{e}-P_\mathrm{net}$): see Figs. \ref{Fig5}(a) and (b). The inhibition of thermionic emission also results in the significant increase of the emitter temperature, as shown in Fig. \ref{Fig5}(c). However, reducing the inter-electrode vacuum gap into the sub-micron regime does not improve PETE performance as well. Although electrons may travel more freely in a sub-micron vacuum space due to the image charge effect and the suppression of the space charge effect, near-field radiative heat transfer increases with the decreasing vacuum gap distance: see Fig. \ref{Fig5}(b). As a result, the emitter temperature cools down for a sub-micron vacuum gap, which ultimately reduces the PETE power output. The optimal gap distance for the best PETE power generation is determined at 2 $\mu$m for the present study, which produces the power density of 1.61 W/cm$^2$ with 17.9 \% energy conversion efficiency for 100$\times$ solar concentration. Fig. \ref{Fig5}(c) also shows the decreasing trend of the photon-enhancement ratio ($n/n_\mathrm{eq}$) as a function of the inter-electrode gap distance. The dependence of $n_\mathrm{eq}$ on the emitter temperature is expressed as $n_\mathrm{eq}=N_C\exp{\left[-(E_g - E_F)/k_B T_\mathrm{E}\right]}$ with $N_C=2\left(m_e^* k_BT_\mathrm{E}/2\pi\hbar^2\right)^{3/2}$ being the effective density of states in the conduction band of Si \cite{sze2006physics}. When the gap distance increases to $d=10$ $\mu$m (i.e., $T_\mathrm{E}=1119$ K), thermally excited electrons ($n_\mathrm{eq}$) greatly outnumber photoexcited electrons ($n-n_\mathrm{eq}$) in the conduction band to make the enhancement ratio close to the unity ($n \approx n_\mathrm{eq}$).

\begin{figure}%[H]
\centering
\includegraphics[width=0.8\linewidth]{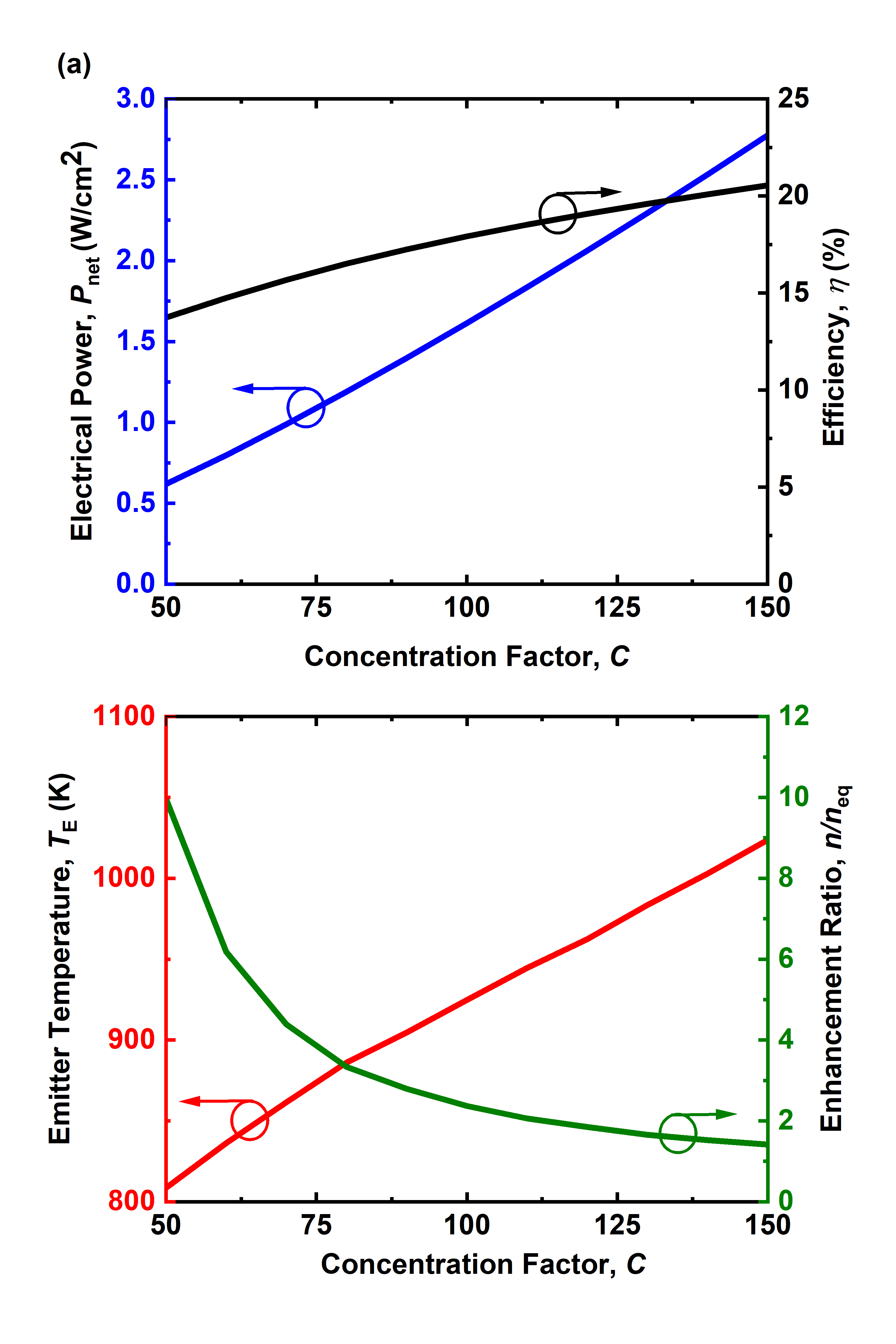}
\caption{}
\label{Fig6}
\end{figure}

Fig. \ref{Fig6} presents the effect of the solar concentration factor onto the PETE performance. As expected, both the PETE power output and the efficiency monotonically increase with the increasing solar concentration factor. However, a high concentration factor drastically decreases the photon-enhancement ratio, indicating that the thermal excitation of electrons should play a more dominant role than photoexcitation in thermionic power generation for high solar concentration. The transition of the electron excitation mechanism is supported by the monotonic increase of the emitter temperature plotted in Fig. \ref{Fig6}(b). Although the photoexcitation rate of electrons in the Si emitter is proportional to the solar concentration factor, most of the absorbed solar energy is converted to heat to increase the emitter temperature. At 150$\times$ suns, for example, the enhancement ratio decreases to $\sim$1.5 while the emitter temperature exceeds 1000 K, suggesting that PETE may be a better energy conversion process for low to mid-scale CSP (e.g., $C\lesssim100$). Unstable semiconducting behaviors at high temperatures (e.g., $\gtrsim1000$ K for Si) are another potential issue of implementing a PETE device for high CSP. It should be noted that the obtained enhancement ratio is much smaller than the previous work \cite{WANG2019}, which predicted the enhancement ratio in the range of several thousand at 100$\times$ suns. We believe that this discrepancy may be due to the aforementioned exponential dependence of $n_\mathrm{eq}$ on the bandgap and temperature of the emitter. While the present study considers the temperature-dependent narrowing of the Si bandgap, which is reduced from 1.12 eV at 300 K to 0.88 eV at 1000 K, \citet{WANG2019} used a constant bandgap (1.4 eV) for calculation. In addition, they did not consider the Auger and Shockley-Read-Hall recombinations, which may lead to an inaccurate prediction of $n$. 

\begin{figure}%[H]
\centering
\includegraphics[width=0.8\linewidth]{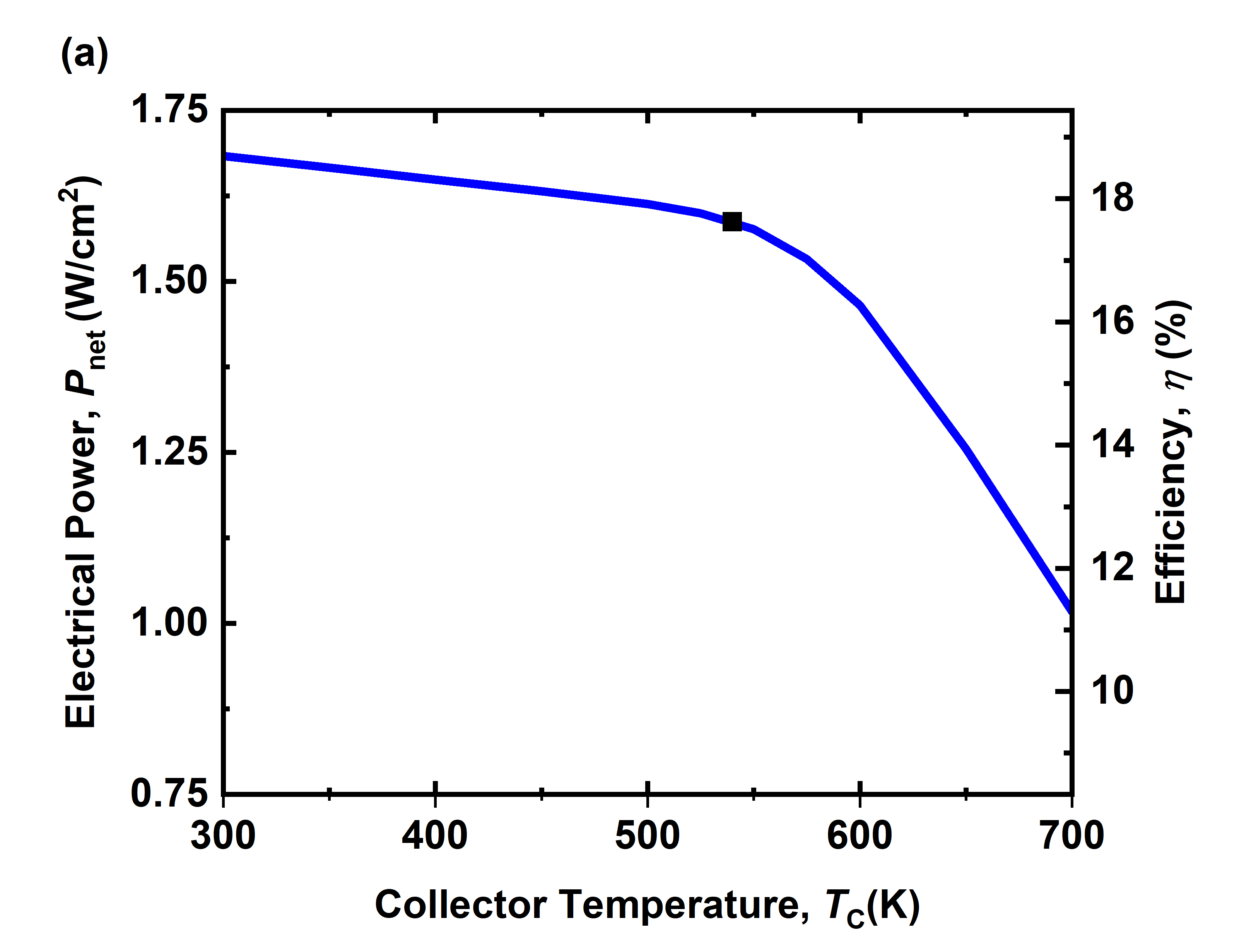}
\caption{}
\label{Fig7}
\end{figure}

Although the collector temperature is set to 540 K to meet the optimal operating condition in the present study, Fig. \ref{Fig7} presents the effect of the collector temperature onto the PETE performance when all other parameters are optimized as given in Table \ref{table:1}. As the collector temperature increases, the PETE performance deteriorates as a result of the increase of the back emission from the collector to the emitter. The back emission becomes more significant when the collector temperature surpasses $\sim$550 K, leading to the drop of the power output and efficiency. Although it seems beneficial to cool the collector below the optimal temperature, it should be noted that lowering the collector temperature requires additional cooling power that is not considered for the calculation of Fig. \ref{Fig7}. The slight gain of the PETE power output by cooling the collector would not be sufficient enough to cover the required cooling load to maintain the collector temperature. 

Table \ref{table:2} summarizes the distribution of incident solar energy into the power output and various losses under the determined geometric configuration and operating conditions as tabulated in Table \ref{table:1}. A portion of incident solar energy that is not absorbed in the device (17.3 \%) accounts for the optical loss. This loss is mostly due to the reflection of solar radiation at the top surface of the glass substrate. Another substrate-related loss is the far-field radiative heat loss to the environment, which takes 27.3 \% of the incoming solar energy. These two loss factors emphasize the importance of substrate material for PETE power generation. Since a substrate is an essential structure that supports a thin semiconducting emitter layer, a substrate material should be carefully selected or engineered to minimize solar reflection in the visible range and thermal emission in the mid-IR range while being transparent in the above-bandgap solar spectrum for effective photoexcitation of electrons in the emitter layer. The optimal design of a substrate material and structure is one of the key requirements to maximize the PETE performance and should be studied in the near future. Thermionic energy transport ($Q_\mathrm{e}$) is the most important mechanism that carries more than 50\% of the incoming solar energy to the collector. Although $\sim$65\% of the thermionic energy is lost as waste heat, this loss is inevitable as far as $P_\mathrm{net}$ is to be maximized: see Eqs. (\ref{Eq:Qe}) and (\ref{Eq:P_net}). The obtained energy conversion efficiency (17.9 \%) is much lower than those predicted in the previous studies \cite{Schwede2010,liu2019effects,WANG2019}. For example, \citet{liu2019effects} predicted the efficiency of 30.2 \% when the PETE device is configured by 0.55$\mu$m inter-electrode gap and operated at $T_\mathrm{E}=1000$ K under 1000$\times$ suns. Their overestimation mainly comes from the fixed temperature assumption without the rigorous energy balance analysis. They also ignored the optical loss and the far-field radiative heat loss to the environment, both of which take a substantial portion of the wasted energy. Assuming the constant emitter bandgap at 1.11 eV without non-radiative recombination processes is another reason of their overprediction of the PETE power output. However, our results show that PETE is still competitive when compared with other CSP technologies \cite{ju2017review,islam2018comprehensive}, and has a room for further improving its performance by implementing novel engineered materials and optimizing the operating conditions.

\section{Conclusion}
\label{S:d}
The present study reports a rigorous energy balance analysis of a PETE device for the evaluation of its performance when the device is to be used for CSP. Instead of assuming ideal materials and operating conditions, we have considered a PETE device configured with real materials (i.e., a boron-doped Si thermionic emitter on top of a glass substrate and a phosphorus-doped CVD diamond collector on top of a tungsten substrate) and accounted for all major energy loss mechanisms occurring in the device under solar irradiation. The dependences of the PETE process on several key parameters, such as the Si emitter thickness, the inter-electrode vacuum gap, the solar concentration factor, and the collector temperature, have been carefully examined to find an optimal design point that would provide the best PETE power output and the energy conversion efficiency. The predicted performance turns out to be not as remarkable as the earlier studies mainly due to the optical and far-field radiative heat losses from the glass substrate, which could be reduced by optically engineering a substrate material and structure to balance the photoexcitation and thermalization processes of electrons in the emitter layer. Although we have considered only one specific PETE device configuration, the developed energy balance analysis and the obtained results can help better understand a realistic feature of the photon-enhanced thermionic energy conversion process for CSP. 

\section*{Acknowledgement}
This research has been supported by National Science Foundation (NSF: ECCS 1611320) and U.S. Department of Energy Solar Energy Technologies Office (DE: EE0008531). The authors greatly acknowledge Dr. Mathieu Francoeur from the University of Utah for fruitful discussions and his valuable comments.



%% References

\clearpage
\bibliographystyle{model1-num-names}
\bibliography{Reference.bib}






% \end{thebibliography}

\clearpage
\begin{table}
\caption{Geometrical and operating parameters used for and optimized by the PETE energy balance analysis.}
\vspace{12pt}
\label{table:1}
%\begin{adjustbox}{width=0.8\textwidth,center}
\centering
\begin{tabular}{lcc}
\hline
Description & Parameters & Values \\
\hline
%The mass of electron&  $m_\mathrm{e}$& $9.11\times 10^{-31} $ $\mathrm{kg}$  \\
Effective mass of electron \cite{Schwede2010}&  $m_e^*$&  $1.0m_e$$^\dag$\\
Effective mass of hole \cite{Schwede2010}&  $m_h^*$&  $0.57m_e$\\
%Richardson constant &$A_\mathrm{E}$,$A_\mathrm{C}$ &$120$ $ \mathrm{A/cm^2K^2}$\\
Emitter doping concentration&  $n_\mathrm{p}$&  $10^{18}$ $\mathrm{cm^{-3}}$\\
Collector doping concentration \cite{Koeck2009}&  $n_\mathrm{n}$&  $10^{18}$
$\mathrm{cm^{-3}}$\\
Emitter electron affinity$^\ddag$&  $\chi$&  $0.7$ $\mathrm{eV}$\\
Collector work function \cite{Koeck2009}&  $\phi_\mathrm{C}$&  $0.9$
$\mathrm{eV}$\\
Ambient temperature&  $T_\infty$&  $300$ $\mathrm{K}$\\
Collector temperature$^\ddag$&  $T_\mathrm{C}$&  $540$ $\mathrm{K}$\\
Concetration factor$^\ddag$&    $C$&    100\\
Glass substrate thickness&  $d_\mathrm{S}$&  $10$ $\mathrm{mm}$\\
Si Emitter thickness$^\ddag$&  $d_\mathrm{E}$&  $50$ $\mathrm{\mu m}$\\
Inter-electrode vacuum gap$^\ddag$&  $d_\mathrm{G}$&  $2$ $\mathrm{\mu m}$\\
Diamond collector thickness \cite{Koeck2006}&  $d_\mathrm{C}$&  $300$ $\mathrm{nm}$\\
%Incident solar radiation&  $Q_\mathrm{inc}$&  $C \times \mathrm{AM1.5D}$\\
\hline
\multicolumn{3}{l}{$^\dag$ $m_\mathrm{e}=9.11\times 10^{-31}$ kg is the electron rest mass.}\\
\multicolumn{3}{l}{$^\ddag$ Based on calculation results}\\
\end{tabular}
%\end{adjustbox}
\end{table}

\clearpage
\begin{table}
\caption{Distribution of solar energy under the operating condition described in Table \ref{table:1}.}
\vspace{12pt}
\label{table:2}
%\begin{adjustbox}{width=0.7\textwidth,center}
\centering
\begin{tabular}{lcc}
\hline
\multirow{2}{*}{Description} & \multicolumn{2}{c}{Distribution} \\ 
                             & (W/cm$^2$)  & (\%) \\
\hline

Optical loss\indent &  1.56 &  17.3 \\
Far-field radiative heat loss\indent &  2.45 &  27.3\\
Near-field radiative heat loss\indent &  0.39 &  4.3\\
Thermionic heat loss\indent &  2.99 &  33.2\\
Electrical power output\indent &  1.61 &  17.9\\
\hline
Solar irradiation (100$\times$ suns)&  9.00 &  100.0\\
\hline
\end{tabular}
%\end{adjustbox}
\end{table}

\clearpage
\begin{center}
\textbf{{Figure Captions}}
\end{center}
\noindent \textbf{Fig. 1:} \indent A schematics of the modeled PETE device consists of a $p$-doped silicon emitter deposited in a glass substrate and separated at a sub-wavelength gap from the diamond thin-film collector. For modeling, we have used 100$\times$ concentration of AM 1.5D as incident solar radiation ($C \times Q_\mathrm{s}$), 10 mm substrate thickness ($d_\mathrm{S}$), $10  \mathrm{\mu m}$ emitter thickness ($d_\mathrm{E}$) and $2  \mathrm{\mu m}$ vacuum gap ($d_\mathrm{G}$), unless otherwise mentioned. Here, we considered constant collector temperature ($T_\mathrm{C}=540\mathrm{K}$), while energy balance dictates the substrate and emitter temperatures. For energy balance, we discretized the substrate into $N$ layers when the temperature of the emitter was considered to be uniform.
\vspace{12pt}

\noindent \textbf{Fig. 2:}(a) \indent The spectral irradiance of 100$\times$ solar energy (AM 1.5D) and corresponding absorption in the emitter and the substrate, respectively. The shaded region refers to the photon energy higher than the energy gap of the silicon at its operating temperature ($T_\mathrm{E} = 925$ K) for the default conditions as summarized in Table \ref{table:1}. Photons in this region photoexcite electrons if absorbed in the Si emitter. (b) The temperature distribution in the emitter structure, where $z=0$ refers to the top surface of the substrate. Part of solar energy absorbed by the emitter structure is converted to heat to increase the emitter temperature. The temperature at the top surface of the glass substrate is lower than that of the Si emitter due to the radiative heat loss to the surroundings. 
\vspace{12pt}

\noindent \textbf{Fig. 3:} \indent (a)The net current density, (b) the net electrical power and efficiency, and (c) the emitter temperature as a function of the operating voltage for the electron affinity of 0.6,0.7,0.8 and 0.9 eV. A Si emitter with a higher electron affinity generates a lower current density and power output and a higher emitter temperature for the operating voltage below $\sim$0.6 V. However, these curves are merged as the operating voltage further increases to the Boltzmann regime. Here, the operating conditions are the same as described in Table \ref{table:1}, except varying electron affinities.
\vspace{12pt}

\noindent \textbf{Fig. 4:} \indent (a) Solar energy absorption in the Si emitter and the glass substrate as a function of the Si emitter thickness and (b) the resultant net electrical power and efficiency. The solar absorption is saturated at silicon thickness around $50 \mu$m, where $\sim$57 \% of the incident solar energy is absorbed by Si and $\sim$26 \% is absorbed by glass. The rest of the incident solar energy is mostly reflected from the top surface of the glass substrate. The electrical power increases with silicon thickness up to $50 \mu$m  owing to the increased solar absorption. However, it starts decreasing with a further increase of the Si emitter thickness due to a higher recombination loss.
\vspace{12pt}

\noindent \textbf{Fig. 5:} \indent (a)The net electrical power and efficiency, (b) the heat losses, and (c) the emitter temperature and enhancement ratio as a function of the inter-electrode vacuum gap. At a lower gap, near-field radiation loss gives a cooling effect, which decreases the temperature and electrical power. Conversely, at a higher gap, the space charge effect increases the potential barrier, which leads to lower electric power and higher temperature. For our PETE device configuration, $2 \mu$m is determined as the optimum gap distance as marked with a sold square point in (a). The increasing emitter temperature with a gap lowers the enhancement ratio as electrons are more thermally excited than by photoexcitation. 
\vspace{12pt}

\noindent \textbf{Fig. 6:} \indent (a) The net electrical power and efficiency, and (b) the emitter temperature and enhancement ratio as a function of solar concentration factor. Both the power and efficiency keep increasing with the increasing concentration factor. The emitter temperature also monotonically increases as the concentration factor increases. As a result, the enhancement ratio decreases due to the dominant thermal excitation of electrons. 
\vspace{12pt}

\noindent \textbf{Fig. 7:} \indent The net electrical power and efficiency as a function of the collector temperature. As the collector temperature increases above 540 K, the performance of the PETE device plummets due to the back emission from the collector. Although the obtained result shows a slightly better PETE performance at lower collector temperatures, it should be noted that the calculation does not consider a cooling load of the collector.

\clearpage

\clearpage

\clearpage

\clearpage

\clearpage

\clearpage

\clearpage

\end{document}